\documentclass[openacc]{rstransa}

\newcommand{\skyday}{sky$^{-1}$\,day$^{-1}$}
\newcommand{\dtfid}{$\Delta t_{\text{fid}}$}
\newcommand{\erad}{$\theta_\text{E}$}

\def\refbf#1{#1}

\titlehead{Review}

\begin{document}

\title{Fast Radio Bursts and the radio perspective on multi-messenger gravitational lensing}

\author{
In\'es Pastor-Marazuela$^{1,2}$}

\address{$^{1}$Jodrell Bank Centre for Astrophysics, University of Manchester, Oxford Road, Manchester M13 9PL, UK\\
$^{2}$Rubicon Research Fellow}

\subject{cosmology, gravitational lensing}


\corres{\email{ines.pastor.marazuela@gmail.com}}

\begin{abstract}
Fast Radio Bursts (FRBs) are extragalactic millisecond-duration radio transients whose nature remains unknown. The advent of numerous facilities conducting dedicated FRB searches has dramatically revolutionised the field: hundreds of new bursts have been detected, and some are now known to repeat. 
Using interferometry, it is now possible to localise FRBs to their host galaxies, opening up new avenues for using FRBs as astrophysical probes. One promising application is studying gravitationally lensed FRBs.
This review outlines the requirements for identifying a lensed FRB, taking into account their propagation effects and the importance of capturing the amplitude and phase of the signal. It also explores the different lens masses that could be probed with FRBs throughout the duration of an FRB survey, from stellar masses to individual galaxies. This highlights the unique cosmological applications of gravitationally lensed FRBs, including measurements of the Hubble constant and the compact object content of dark matter. Finally, we discuss future radio interferometers and the prospects for finding gravitationally lensed FRBs.
\end{abstract}

\begin{fmtext}
\section{Fast Radio Bursts: Introduction}

The astrophysical phenomena known as Fast Radio Bursts (FRBs) are extragalactic radio transients with millisecond durations. Produced at \refbf{distances up to cosmological scales}, they can reach energies as high as $10^{44}$\,erg, and they occur at a rate of $\gtrsim1000$\,\skyday\ for typical luminosities \citep{petroff_fast_2019, petroff_fast_2022}.
Since their discovery in 2007 \citep{lorimer_bright_2007}, hundreds of FRB discoveries have been published. 
The majority of these bursts have been seen just once; these are known as one-offs or non-repeaters \citep{chimefrb_collaboration_first_2021}.
\end{fmtext}

\maketitle

However, about 50 FRBs have been found to repeat \citep{spitler_repeating_2016, chimefrb_collaboration_chimefrb_2023}\footnote{To see the full catalogue of FRBs, visit the Transient Name Server (TNS): \url{https://www.wis-tns.org}}.
Unlike pulsars, the wait time between consecutive bursts does not appear to show a periodic pattern \citep[e.g.,][]{niu_fast_2022}, but two highly active repeaters have revealed periodic activity cycles of 16 days \citep{chimefrb_collaboration_periodic_2020, pastor-marazuela_chromatic_2021} and 160 days \citep{rajwade_possible_2020}. It is not known yet whether one-offs and repeaters have the same origin or if they are instead produced by different sources or mechanisms, however, their morphologies --spectro-temporal structure \refbf{after correcting for dispersion}-- provide valuable clues; while one-offs can be either broadband, narrowband, or have multiple temporal components peaking at the same frequency, repeaters often display multiple components that drift downwards in frequency --referred to as the `sad trombone effect'-- \citep{pleunis_lofar_2021, hessels_frb_2019}.

The cosmological nature of FRBs was initially inferred through their dispersion measures (DMs). Dispersion occurs when the radio waves travel through an ionised plasma; this results in a frequency-dependent refraction index, which results in a larger delay in the arrival time for lower frequencies, which can be used to estimate the distance to the emitting source. The first FRBs were detected with DMs far in excess of the expected contribution from the Milky Way (MW) and its halo, and at high galactic latitudes, thus suggesting an extragalactic origin \citep{thornton_population_2013}. 
Direct confirmation came through the localisation of the first repeating FRB to a dwarf galaxy at a redshift of 0.19 in 2017 \citep{chatterjee_direct_2017, tendulkar_host_2017}. Since then, more than 40 FRBs have been localised to their host galaxies. Although repetition allows for follow up observations and the use of Very Long Baseline Interferometry (VLBI) techniques to localise FRBs with milli-arcsecond accuracy \citep{marcote_repeating_2020, nimmo_milliarcsecond_2022}, one-off FRBs seem more abundant. Current FRB surveys can achieve (sub-)arcsecond precision in the localisation of single bursts, which is often sufficient to unambiguously identify their host galaxies. About two thirds of the localised FRBs are thus one-offs \citep{bannister_single_2019, law_deep_2024, rajwade_study_2024}. These localised FRBs span distances from 3.6\,Mpc \citep{kirsten_repeating_2022} up to a redshift of 1 \citep{ryder_luminous_2023}, although from the observed DMs some FRBs could originate from redshifts up to 3 and beyond \citep{pastor-marazuela_comprehensive_2024}.

Many models have been proposed to explain the origin of FRBs \citep{platts_living_2019}, but in spite of the diverse breakthroughs on the FRB properties since they were first discovered, their nature and emission mechanism remain unknown. Nevertheless, the detection of an extremely bright radio burst from the Galactic magnetar SGR\,1935+2154 in 2020 bridged the luminosity gap between Galactic radio loud neutron stars and extragalactic FRBs \citep{bochenek_fast_2020, chimefrb_collaboration_bright_2020}, and proved that at least some FRBs must be produced by magnetars. However, we still cannot determine whether all FRBs are magnetars, with younger, active ones being seen to repeat, while older, less active ones are just seen once. Instead, some one-offs could be produced by cataclysmic events, such as compact binary mergers or hyper-massive neutron stars slowing down to become black holes --blitzars \citep{falcke_fast_2014}--, and their lack of repetition wound be an intrinsic property.
 
The study of FRB host galaxy properties could give further clues to the FRB progenitors. Different transient phenomena, which mark the violent transition between consecutive stages in stellar evolution, are known to preferentially occur in specific galaxy types. Comparing FRB host galaxy properties to the hosts of various transients, is showing that, while long gamma-ray bursts (LGRBs) and superluminous supernovae (SLSNe) are produced in hosts with properties differing from those of FRBs, the hosts of core-collapse supernovae (CCSNe) and short gamma-ray-bursts (SGRBs) are indistinguishable from those of FRBs. This might suggest an evolutionary link between CCSNe or SGRBs with FRBs \citep{bhandari_characterizing_2022}. 
Other studies have shown that FRB hosts mostly follow the star-forming main sequence of galaxies \citep{gordon_demographics_2023}, while active galactic nuclei (AGN) are not dominating the population \citep{eftekhari_x-ray_2023}. 

Another way to determine the FRB origin is to study the frequency extent of the emission, and to ascertain the presence of any multi-wavelength and multi-messenger counterparts.
Within the radio, FRBs have been detected in frequency ranges spanning from 110\,MHz \citep{pastor-marazuela_chromatic_2021, pleunis_lofar_2021} up to 8\,GHz \citep{gajjar_highest-frequency_2018}. However, no multi-wavelength or multi-messenger counterparts have been detected yet in spite of numerous searches and follow-up campaigns \citep[e.g.][]{trudu_simultaneous_2023, pearlman_multiwavelength_2023, curtin_constraining_2024, bhardwaj_gw190425_2023}.

The cosmological origin of FRBs makes them invaluable tools to address several cosmological questions. 
One of the best studied applications combines the use of well localised FRBs with known host galaxy redshifts and their DMs. This has been used to measure the baryonic content of the Intergalactic Medium (IGM), which is otherwise challenging to detect, leading to the so-called Macquart relation, which relates the DM to the source redshift \citep{macquart_census_2020}.

With FRBs becoming ever more abundant, and new instruments being built or improved to increase their detection rate and localisation accuracy, FRBs will soon become powerful tools to address some current astrophysical and cosmological questions from a novel perspective, including the study of gravitationally lensed FRBs. This work reviews the current research on gravitationally lensed FRBs, the techniques and observables that can be used to detect them, the challenges that one might encounter, the future prospects for detecting them with upcoming instruments, and the potential synergies with multi-wavelength and multi-messenger FRB counterparts.

\section{Gravitationally lensed FRBs}

\begin{figure}[!h]
\centering\includegraphics[width=0.8
\textwidth]{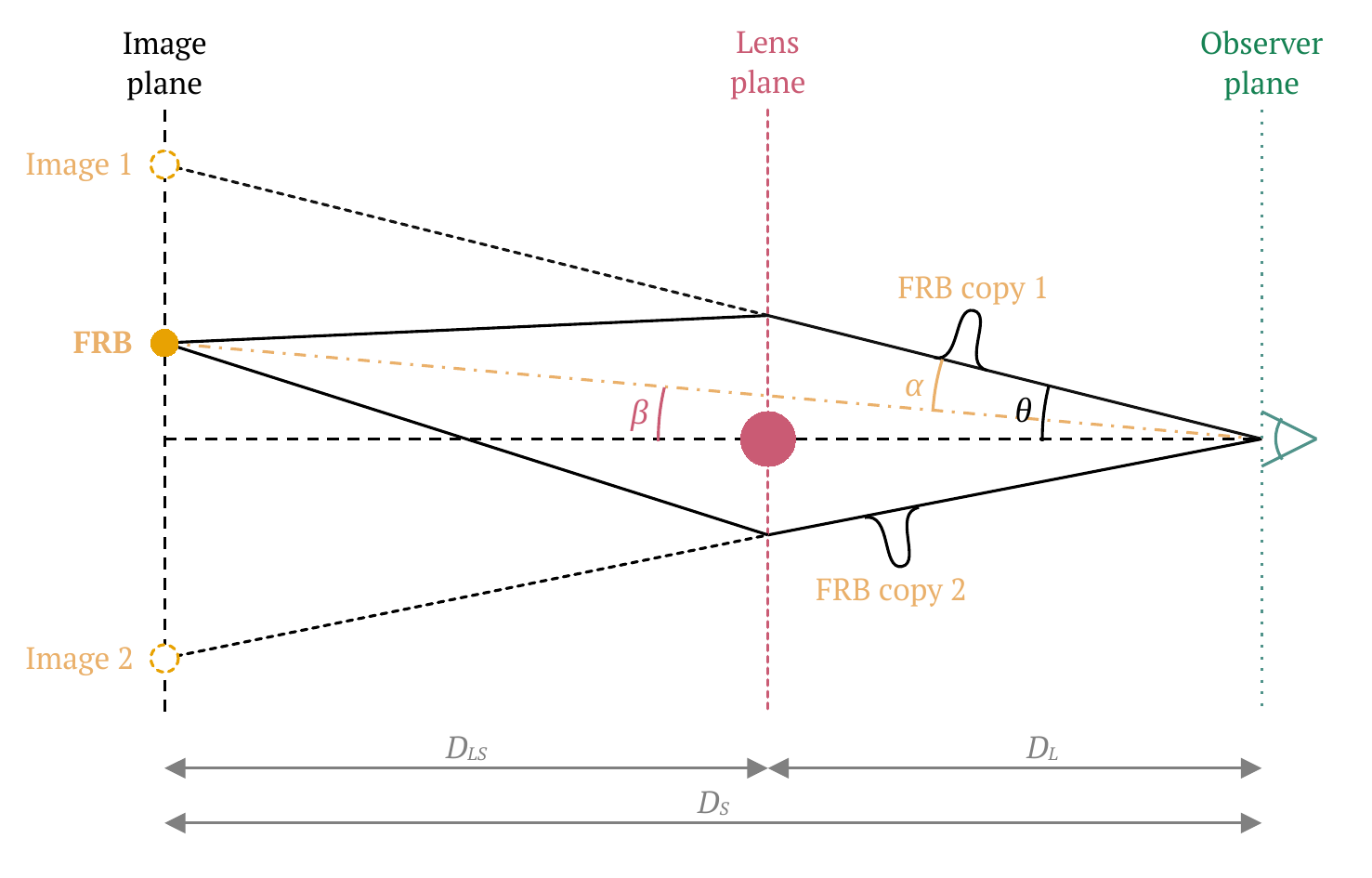}
\caption{Diagram of a gravitationally lensed FRB.}
\label{fig:gl_diagram}
\end{figure}

Gravitational lensing is produced when the light rays of an astrophysical source are deflected by a matter distribution located between the source and the observer. If the deflection angle is sufficiently large, multiple images of the same source can be distinguished. The different travel paths will also result in different time delays, which can be measured using transient and variable sources.
In the analysis of gravitationally lensed sources, the mass distribution throughout the line of sight of a source might produce complex deflection patterns, but it is often assumed that the deflection will be dominated by a single thin lens along the propagation path, as illustrated in Fig.~\ref{fig:gl_diagram}.
The short duration of FRBs enable the measurement of millisecond, and even microsecond time delays $\Delta t$, while their localisation accuracy could resolve images with angular separations $\Delta\theta$ of the order of arcseconds. 

However, two or more lensed FRB copies travelling through divergent travel paths are likely to experience distinct propagation effects, which will modify differently the observed lensed copies and could complicate their identification \refbf{\citep{leung_constraining_2022}.}
On top of the dispersion, to which the gravitational lens could contribute differently depending on the propagation angle, the lensed FRB images could experience different scattering and scintillation.
Scattering is produced when the burst travels through an inhomogeneous medium that scatters the waves in multiple directions, and the resulting variable path lengths will be observed as an exponential decay of the burst intensity as a function of time.
Scintillation is also produced when the burst travels through an inhomogeneous medium. In this case, the effect results in patchy intensity variations with frequency as the waves coming from different directions interfere constructively or destructively \citep[][chapter 4]{lorimer_handbook_2004}.
On Earth, Radio Frequency Interference (RFI) can produce a lot of noise, reduce the effective bandwidth, and impede the FRB detection. 
These effects must be taken into account when attempting to identify lensed FRB copies, but as detailed below, there are techniques to overcome them.

\subsection{Identifying lensed FRBs}

Three main techniques can be used to identify gravitationally lensed FRBs; auto-correlating voltage data for small time delays, cross-correlating intrinsic FRB structure and polarisation for longer time delays, and using their precise localisation. These techniques are all described below.

\subsubsection{Voltage data}

Two lensed copies of the same FRB should preserve the same phase of the electric field when the time delay is $\lesssim1$\,s. Several FRB surveys with real-time detection pipelines are now able to trigger a raw voltage data capture of a new burst upon its detection \citep{cho_spectropolarimetric_2020, law_deep_2024, chimefrb_collaboration_updating_2023, rajwade_study_2024}, opening up the possibility of cross-correlating the FRB electric field phases to identify lensed copies with time delays within the duration of the trigger. These voltage data triggers often contain a few hundred milliseconds of data, and thus the observed time delays would be within the microlensing regime.
In \citep{kader_high-time_2022}, the authors developed a technique to identify lensed FRBs in voltage data from the CHIME/FRB instrument, where they can detect time delays from the minimal time resolution of 1.25\,ns up to the trigger duration of 100\,ms.
Such time delays would correspond to a gravitational lens with a mass up to a few hundred solar masses. In such cases, the propagation path would be very similar and the difference in propagation effects is likely to be negligible.

\subsubsection{FRB intrinsic structure}

Lensing by larger masses will produce time delays beyond the duration of a single voltage dump. The FRB lensed copies are also likely to experience different propagation effects that will cause decoherence of the electric field phases \citep{leung_constraining_2022}. In such cases, the intrinsic structure and polarisation of the bursts can be used to identify lensed FRBs. 

On top of having access to the phase information, voltage data allows to achieve a much higher time resolution than that typically used for FRB searches. This time resolution can be as short as microseconds and it is starting to reveal that complex spectro-temporal structures are prevalent among FRBs, both one-offs and repeaters \citep{pastor-marazuela_comprehensive_2024, faber_morphologies_2023, nimmo_burst_2022}.

Furthermore, the voltage data often allows one to measure the polarisation properties of the FRBs. Intrinsic polarisation properties include the linear and circular polarisation fractions, as well as the polarisation position angles. Linearly polarised radio waves travelling through an ionised plasma with a magnetic field parallel to the propagation direction undergo Faraday rotation, an additional propagation effect observed as a phase wrapping of Stokes Q and U as a function of frequency \citep[][chapter 4]{lorimer_handbook_2004}. This propagation effect can be quantified through the Rotation Measure (RM) and subsequently corrected \citep[e.g.][]{mckinven_polarization_2021} to reveal the intrinsic polarisation properties. Within the FRB population, these polarisation properties have been observed to vary widely, even within the same burst subcomponents \citep{pandhi_polarization_2024}. This has already been used to discriminate bursts with several intrinsic components from being gravitationally lensed events \citep{cho_spectropolarimetric_2020}, and could further be used to identify strongly lensed FRBs in the future. 

Hence, cross correlating the intrinsic FRB structure and polarisation properties after correcting for propagation effects like dispersion, scattering, and Faraday rotation could allow for the identification of strongly lensed events with time delays from seconds to months, and even years.

\subsubsection{Sub-arcsecond localisation}

Since the identification of the first FRB host galaxy in 2017 \citep{chatterjee_direct_2017}, several FRB surveys with detection rates of a few tens per year have developed the required software to achieve routine localisations of a large fraction of the detected one-off bursts \citep{macquart_census_2020, law_deep_2024, rajwade_study_2024}. Currently, CHIME/FRB, the leader in terms of FRB detection rates, can achieve sub-arcminute precision for bursts with voltage data, which is comparable to the localisation accuracy of Swift-BAT\footnote{Swift GRB table: \url{https://swift.gsfc.nasa.gov/archive/grb_table/}} in GRB searches, and much more accurate than current Gravitational Wave (GW) localisation \citep[tens to thousands of square degrees,][]{abbott_gwtc-1_2019}. The inclusion of the CHIME/FRB outrigger stations \citep{lanman_chimefrb_2024} will soon result in milliarcsecond localisations for hundreds of bursts.

An accurate transient localisation allows one to determine whether its host galaxy is strongly lensed. This could be used to determine whether two or more detected bursts are lensed copies of the same event or instead different bursts altogether. Alternatively, one could predict if and when a lensed copy is expected to arrive. The distortion of the host galaxy can give access to the lens properties, like mass and geometry, and thus provide with information to estimate when the lensed copy will arrive --or if it already arrived in the past and was missed. If two or more copies of a strongly lensed FRB are detected, the absolute time delays can be determined with (sub)millisecond precision, which makes strongly gravitationally lensed FRBs excellent tools for several cosmological applications, as will be discussed in Section~\ref{sec:applications}.

\subsection{Observable lens masses}

The identification techniques listed above give access to a broad range of time delays between multiple images. However, several factors limit the range of observable lens masses that one can study with FRBs. These are mainly the time resolution of the instruments, going from milliseconds for searches using the intrinsic structure, to microseconds when using the phase information. On the other end, the maximal time delays that can be observed are limited by the duration of the FRB surveys, typically between two and six years and not continuous on a given field \refbf{\citep[e.g.][]{pastor-marazuela_comprehensive_2024, chimefrb_collaboration_first_2021}}. 
We can compare these observable delays to the fiducial time delay \dtfid, which gives the arrival time difference between an image deflected by $\theta_{\text{E}}$ and the unlensed line of sight. For a point lens mass, it is given by the following expression, as derived in \citep[Eq.~21,][]{oguri_strong_2019}:
\begin{equation} \label{eq:fiducial_timescale}
    \Delta t_{\text{fid}} \sim 1.97\times 10^{-5} \text{s} \times (1+z_l)\left( \dfrac{M}{M_\odot}\right),
\end{equation}
where $z_l$ and $M$ are respectively the redshift and the mass of the lens, for a point lens mass. This can also be applied to different lens mass distributions by replacing $M$ by $M(<\theta_{\text{E}})$.

\begin{figure}[!h]
\centering\includegraphics[width=0.8\textwidth]{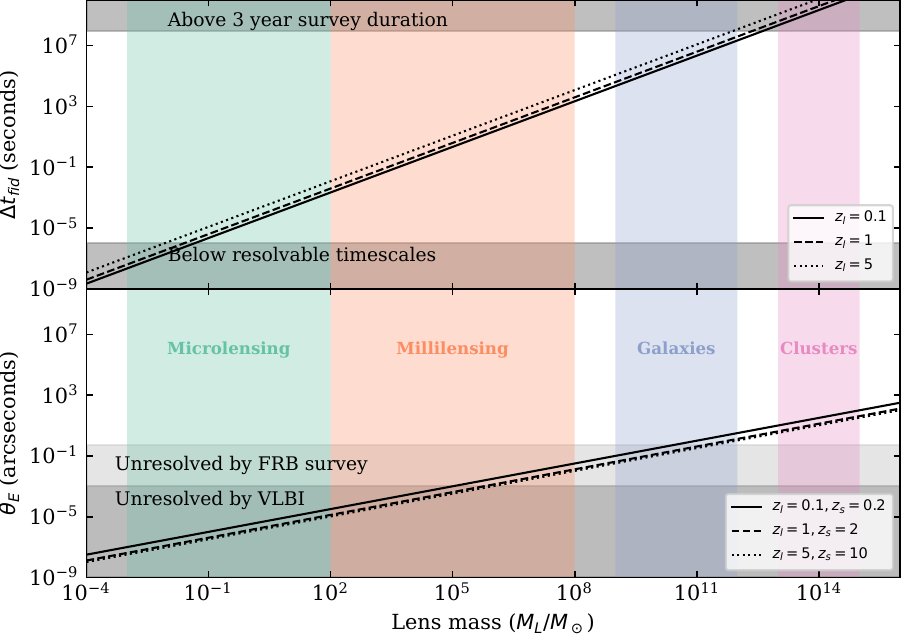}
\caption{Observable lens masses through gravitationally lensed FRBs. The top panel shows the fiducial timescale \dtfid\ as a function of lens masses for lenses at redshifts $z_l=(0.1, 1, 5)$; the bottom panel shows the Einstein radius \erad\ for the same lens masses, $z_l=(0.1, 1, 5)$, and $z_s=(0.2, 2, 10)$. The coloured shaded regions indicate different lensing regimes; green shows microlensing (MACHOs, PBHs, stars, free floating planets), orange shows millilensing (Intermediate mass black holes, dark matter halos), purple shows lensing by galaxies, and pink lensing by clusters. The grey shaded areas show the regions that cannot be identified with current or upcoming surveys. Adapted from \citep[][Fig.~2]{connor_stellar_2023}.}
\label{fig:lens_masses}
\end{figure}

The angular resolution limits the separation between the images that can be resolved. 
The typical angular distances between gravitationally lensed images are given by the Einstein radius, which is given by the following expression for point-like lenses; 
\begin{equation}
    \theta_{\text{E}} = \sqrt{\dfrac{4GM}{c^2} \dfrac{D_{LS}}{D_L D_S}}.
\end{equation}
The ranges of observable time delays, angular separations, and consequent lens masses are shown in Fig~\ref{fig:lens_masses} \citep[adapted from][]{connor_stellar_2023}. As can be seen in the figure, the observable time delays are able to resolve lens masses from $\sim10^{-2} M_\odot$ up to $\sim10^{12} M_\odot$. This goes from the microlensing regime up to lensing by massive galaxies. 
The angular resolution of current radio interferometers can resolve lensing by galaxies or clusters. VLBI techniques could resolve down to the millilensing regime, but currently, this localisation technique can only be applied to repeating FRBs.
Hence, while time delays can be detected in a broad range of scales, the angular separation between the lensed images will only be resolved for galaxy lenses and above.

\subsection{Detection probability}

The detection probability of a gravitationally lensed FRB depends on several factors, which can be broken down into the likelihood that an FRB encounters a gravitational lens along its propagation path, the probability that the FRB will be bright enough to be detected with current instruments, and the probability that the instruments are observing at the time when the FRB reaches Earth.

The probability of encountering massive objects acting as gravitational lenses along the propagation path is steeply dependent on redshift, and thus higher redshift FRBs will have a greater likelihood of being lensed. The optical depth for gravitational lensing $\tau(z_s)$, is given by \citep[][Eq.~4]{connor_stellar_2023}:
\begin{equation} \label{eq:optical_depth}
    \tau(z_s) = \int^{z_s}_0 dz_l \int n(\sigma, z_l) \sigma \dfrac{d^2 V}{d\Omega dz_l},
\end{equation}
where $n(\sigma, z_l)$ is the number density of gravitational lenses along the propagation path at redshifts $z_l$ with cross sections $\sigma \sim \pi\theta_E^2$, and $\frac{d^2 V}{d\Omega dz_l}$ is the comoving volume element per steradian per redshift. Since we are in the optically thin regime, the lensing probability can be approximated as $P_{\text{lens}, z_s}=1-e^{-\tau}\approx\tau(z_s)$ \citep{munoz_lensing_2016}. 
Since the Einstein radius depends on the mass distribution of the lens, the optical depth will have different expressions for a point mass and for the mass distribution in galaxies or clusters \citep[][and references therein]{oguri_strong_2019, connor_stellar_2023, munoz_lensing_2016}.
As shown \refbf{in} the left panel of Fig.~\ref{fig:surveys}, strong lensing from massive galaxies reaches $\tau(3\leq z_s\leq 5)\sim10^{-3}$, while lensing by dark matter compact halos is highly dependent on the fraction of those; if a fraction of $10^{-3}$ of dark matter is made of compact halos, we expect $\tau(3\leq z_s\leq 5)\sim10^{-3}$, while if that fraction is $10^{-4}$, $\tau(3\leq z_s\leq 5)$ decreases to $10^{-4}$. At the typical FRB redshifts $z_s\sim0.5$, the fraction of expected gravitationally lensed FRBs along their propagation path is $\sim10^{-5}$.

We can use the DM distribution from large FRB samples to infer their intrinsic redshift distribution \citep{chimefrb_collaboration_first_2021}. This observed redshift distribution depends on the sensitivity and observing setup of the instrument,  on the intrinsic luminosity function of the FRB population, as well as on the intrinsic FRB redshift distribution. The current CHIME/FRB observed population peaks at redshift $\sim0.5$, and only $\sim3\%$ of the observed bursts have an inferred redshift $>2$. Although more sensitive instruments have observed FRBs with inferred redshifts $\gtrsim3$ \citep{pastor-marazuela_comprehensive_2024}, the detection rate of those is currently too low to expect a gravitationally lensed FRB to be detected.

Although many FRBs might have intrinsic fluxes that fall below the detection limits, magnification will play a role in increasing the observed flux of a lensed event, and boost the probability of detecting gravitationally lensed FRBs \citep{oguri_strong_2019}. This is quantified with a magnification bias factor that multiplies Equation~\ref{eq:optical_depth}. This in turn could be used as a technique to detect the most distant FRBs, which has applications to study the FRB rate evolution compared to the Star Formation History of the Universe.

Lastly, survey observing strategies yield a time delay selection function. While strong lensing of galaxies can be identified with a single image, finding gravitationally lensed copies of the same transient requires observing the same region of the sky when each copy arrives \citep{connor_stellar_2023}. Some surveys like CHIME/FRB have large Fields of View (FoV), but are constantly pointing in the same direction (the zenith), while other surveys have smaller FoVs but can point in the desired direction while a target is visible in the sky.
Detecting a lensed event will thus require sensitive instruments with large FoVs, or survey strategies designed to detect these lensed transients, for instance by spending the majority of their time pointing in the same directions of the sky. 

\section{Applications and case studies} \label{sec:applications}

Below we discuss some of the studies that have proposed to use gravitationally lensed FRBs and the questions that could be addressed with these.

\subsection{Microlensing: searching for compact halo objects}

The presence of dark matter in the Universe has been used to interpret several gravitational effects in large scale structures that cannot otherwise be explained by general relativity \citep{trimble_existence_1987}. However, the nature of dark matter remains a mystery after decades of research in the field. One of the first candidates suggested to explain the excess mass were compact objects residing in the halos of galaxies and clusters, such as Massive Compact Halo Objects (MACHOs) or Primordial Black Holes (PBHs). Searches for these objects have been carried out primarily through the study of gravitational microlensing \citep{aubourg_search_1999, alcock_macho_2001, wyrzykowski_ogle_2011}, setting strong constraints on the fraction of lower mass objects ($\lesssim5\%$ in the mass range $10^{-7}-30M_{\odot}$), while the study of binary systems with wide separations in the MW halo constrains the fraction of massive objects \citep[$\lesssim40\%$ for $\gtrsim100M_{\odot}$,][]{quinn_reported_2009}.

Gravitationally lensed FRBs could provide constraints on the mass range between $30M_\odot$ and $100M_\odot$. While the angular separation between lensed images would be too small to be resolved with current instruments, the time delay between lensed copies would be sufficient to characterise the lensing object and the fraction of compact halo objects that form dark matter \citep{munoz_lensing_2016, sammons_first_2020}. 

A recent search for gravitationally lensed FRBs in 172 bursts with voltage data from CHIME/FRB found no lensed events, and this was used to place constraints on the fraction of dark matter made of compact objects with masses $\sim10^{-3}M_{\odot}$ to be $f\lesssim80\%$ \citep{leung_constraining_2022}. 
The number of FRBs used in this study is still low; a non-detection of lensed events in $10^4$ FRBs could constrain that fraction to $f\lesssim0.9\%$ for masses $\sim30M_{\odot}$ \citep[][in this study the voltage \refbf{data are} not taken into account, and thus the constrained mass ranges could be extended to lower masses]{munoz_lensing_2016}. Such constraints would greatly improve the current limits set by dark matter experiments \citep{aubourg_search_1999, alcock_macho_2001, quinn_reported_2009}, and the number of required FRBs could be achieved in the coming decade.

\subsection{Measuring the Hubble constant}

The Hubble constant, $H_0$, is a fundamental cosmological parameter to measure the expansion rate of the Universe, test cosmological models, and determine extragalactic distances. However, some of the different techniques that have been applied to measure $H_0$ have yielded disparate results, leading to the `Hubble tension'. Low redshift measurements using the `cosmic ladder' find $H_0=(74.03\pm1.42)$\,km\,s$^{-1}$\,Mpc$^{-1}$ \citep[SH0ES programme,][]{riess_large_2019}, while early Universe measurements using the Cosmic Microwave Background (CMB) anisotropies observed with the \textit{Planck} mission yield $H_0=(67.4\pm0.5)$\,km\,s$^{-1}$\,Mpc$^{-1}$ \citep[][]{aghanim_planck_2020}. These values are discrepant at a $4.4\sigma$ significance level. If the difference is not due to poorly understood systematic errors, it could reveal new physics beyond the standard $\Lambda$CDM model that we do not understand yet. These could be for instance modified theories of gravity, or a poorly understood dark energy equation of state.

To solve the Hubble tension, it is thus important to use different techniques to measure $H_0$. Using multi-messenger compact object mergers provides new ways of independently determining $H_0$, and it was already tested with GW170817 and its associated kilonova \citep{abbott_gravitational-wave_2017}. 
FRBs localised to their host galaxies have recently been used to measure $H_0$ from the dispersion measure-redshift relation \citep{hagstotz_new_2022}. However, this method relies on assumptions about the DM contribution at the host galaxy and the Milky Way, which are often hard to constrain. 
Time-delay cosmography, an additional technique to measure $H_0$, leverages the time delay between gravitationally lensed images to measure $H_0$ \citep{treu_time_2016}, and it has already been applied to strongly lensed quasars \citep[H0LiCOW,][]{wong_h0licow_2020}. Lensed quasars are common enough that the lensing time delay can be measured in a sufficiently large number of sources. However, the long timescales over which they vary results in timing uncertainties of several days (1 to 6\% relative errors). 

These large uncertainties could be overcome with the use of FRBs. Due to their short duration, the lensed time delays can be measured with micro/millisecond accuracy, or a precision of ${\sim10^{-9}-10^{-12}}$ for galaxy-lensing delays spanning several days to weeks. Additionally, the scarcity of \refbf{AGN} among FRB host galaxies \citep{eftekhari_x-ray_2023} allows for better modelling of the lens parameters.
\refbf{While the DM contribution from the lens may vary across lensed images travelling through different propagation paths, the FRB arrival time extrapolation at an infinite frequency ensures that time delays are not affected by these DM differences. Moreover, the different DMs between lensed images could provide additional insights into the matter distribution near the lens, offering a complementary tool to constrain both the galaxy profile and $H_0$ when combined with other wavelengths or observables.}

Previous studies have primarily focused on repeating FRBs due to the milli-arcsecond localisation accuracy that can be achieved through VLBI techniques \citep{li_strongly_2018, liu_prospects_2019, wucknitz_cosmology_2021}. 
Furthermore, lensed repeating FRBs offer the opportunity to study the evolution of the lens time delay over time, thereby reducing uncertainties on the lens mass distribution \citep{wucknitz_cosmology_2021}.
However, repeating FRBs appear to be quite rare, constituting only $\lesssim3\%$ of the current FRB population. Moreover, the observed repeater DMs are generally lower than those of one-offs, indicating that they originate from lower redshifts \citep{chimefrb_collaboration_chimefrb_2023}, and making them less likely to be gravitationally lensed. Consequently, detecting a repeating FRB lensed by galaxy masses might only be achieved after identifying $10^5-10^7$ FRBs.
Nonetheless, the localisation accuracy of current and upcoming one-off FRB surveys would suffice to identify the lensed host galaxy, thus enabling the use of gravitationally lensed one-off FRBs to measure $H_0$. 
According to \citep{li_strongly_2018}, time delays from 10 gravitationally lensed FRBs could improve the $H_0$ measurement from lensed quasars by a factor of five, while \citep{liu_prospects_2019} conclude that measuring time delays from 30 gravitationally lensed FRBs would help improving the errors on $H_0$ from other experiments by a factor of 2. Although these values are estimated for repeating FRBs, well-localised one-off FRBs could potentially achieve similar accuracies.

\section{Future radio telescopes and prospects}

FRB surveys are at present achieving astounding results. CHIME/FRB has detected thousands of FRBs, including $\sim50$ repeaters \citep{leung_constraining_2022, chimefrb_collaboration_chimefrb_2023}, and it can localise single bursts to a few tens of arcmin$^2$ error regions. ASKAP \citep{macquart_census_2020}, MeerKAT \citep{rajwade_study_2024}, and DSA-110 \citep{law_deep_2024} are able to localise one-off bursts with sub-arcsecond precision, and all of the above are able to store voltage data of real-time FRB detections. These surveys span different ranges of FoVs and sensitivities, but none of them is able yet of achieving the combined detection rate and localisation accuracy required for the identification and study of gravitationally lensed FRBs. 
However, several radio interferometers scheduled to become operational in the coming decade could achieve the required detection rate and localisation accuracy to implement the case studies described above. 

The Canadian Hydrogen Observatory and Radio-transient Detector \citep[CHORD,][]{vanderlinde_canadian_2019} is planned to be an upgrade of CHIME/FRB in Canada. It will combine a very large FoV, high sensitivity, and very broad bandwidth, and could reach detection rates of thousands of FRBs per year. Additionally, it will include outrigger stations \citep[already being tested for CHIME/FRB,][]{lanman_chimefrb_2024}, thus adding VLBI localisation accuracy for one-off FRB detections.
The Square Kilometre Array (SKA), which is being built in South Africa and Australia, will reach sensitivities unparalleled by any other radio telescope to date. Adding its large FoV and long baselines, it could detect and localise $>10^4$ FRBs per year \citep{schilizzi_square_2010, macquart_fast_2015}. Its sensitivity will allow for the detection of high redshift FRBs, and thus increase the chances of detecting gravitationally lensed bursts.
The Deep Synoptic Array (DSA) 2000 will be an array of 2000 antennas installed in the US that will search for FRBs in a 10.6\,deg$^2$ FoV, and it is expected to detect and localise $10^3-10^4$ FRBs per year \citep{hallinan_dsa-2000_2019}.
The Bustling Universe Radio Survey Telescope in Taiwan (BURSTT) is being built with the goal of detecting bright, nearby FRBs. Although it might not be sensitive enough to detect the most distant FRBs, its extremely large FoV will allow for an easier detection of lensed copies of the same burst. Furthermore, it will count with outrigger stations to localise FRBs with arcsecond accuracy \citep{lin_burstt_2022}.

\begin{figure}[!h]
\centering
\includegraphics[height=0.385\textwidth]{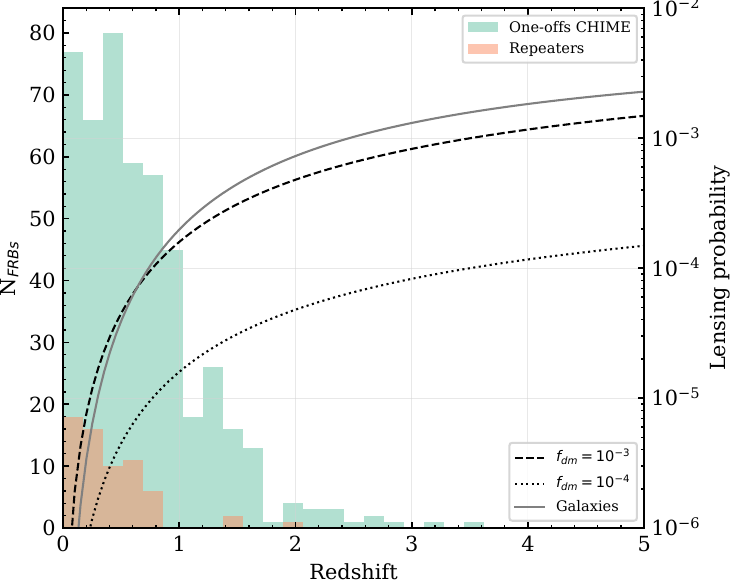}
\hfill
\includegraphics[height=0.385\textwidth]{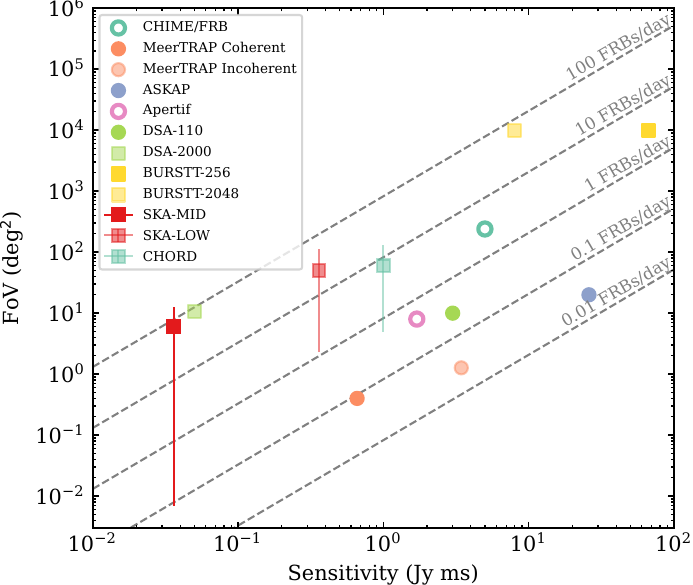}
\caption{\textbf{Left}: Redshift distribution of CHIME one-offs (green) and repeater FRBs (orange), estimated from the Macquart relation. The dashed and dotted lines represent lensing from compact objects if they contribute to a fraction of $10^{-3}$ and $10^{-4}$ of dark matter respectively. The grey solid line shows the probability of lensing by galaxies as a function of redshift. Adapted from \citep{connor_stellar_2023}.
\textbf{Right}: Observing properties of current and upcoming surveys. The field of view of each instrument is plotted as a function of their sensitivity. Circular markers indicate current instruments, while squares show future surveys. Filled markers indicate surveys that can achieve (sub-)arcsecond localisation, while empty markers show those that cannot. Grey dashed lines show the detection rate per 24h of observations.}
\label{fig:surveys}
\end{figure}

The combination of these radio telescopes presents excellent prospects to achieve the FRB detection rates and redshifts required to detect gravitationally lensed bursts. Adding the localisation accuracy that will be reached, gravitationally lensed FRBs are likely to become excellent tools to probe dark matter and improve the constraints on cosmological constants. The properties of the aforementioned surveys can be visualised in Fig.~\ref{fig:surveys}.

Despite the current lack of confirmed multi-messenger/wavelength counterparts to FRBs \citep{pearlman_multiwavelength_2023, curtin_constraining_2024, bhardwaj_gw190425_2023}, future discoveries hold promise for finding them. Magnetars have been observed to emit radio bursts during X-ray or $\gamma$-ray outbursts \citep{bochenek_fast_2020, camilo_transient_2006}. These correlations support FRB models predicting the existence of similar counterparts. If FRBs originate from young neutron stars, it might be possible to observe supernovae or other kinds of explosive transient prior to the onset of the FRB activity. This is supported by the similarity between the host galaxies of CCSNe or SGRBs and FRBs \citep{bhandari_characterizing_2022}.
Alternatively, if some FRBs are genuine non-repeaters, they could originate from cataclysmic events such as compact binary mergers and other explosive transients \citep{falcke_fast_2014, piro_magnetic_2012}. 
As observational techniques and detection sensitivity improve, particularly in X-ray, $\gamma$-ray, and GW astronomy, the identification of these counterparts could significantly advance our understanding of the FRB origins, and open up the field of multi-messenger gravitational lensing using FRBs.

\section{Summary and conclusions}

Although the concept of time-domain gravitational lensing was proposed decades ago \citep{refsdal_possibility_1964}, it has only recently become a reality \citep{kelly_multiple_2015, goobar_iptf16geu_2017, rodney_gravitationally_2021, wong_h0licow_2020}. The rapidly evolving field of FRBs is expected to soon produce the necessary data to search for gravitationally lensed events, and address some of the most pressing questions in astrophysics and cosmology.
Since the time delays between gravitationally lensed copies of FRBs can be measured with micro/millisecond accuracy, through the use of voltage data \citep{kader_high-time_2022} or the intrinsic burst structure, they could provide a new method to measure the Hubble constant $H_0$ and other cosmological parameters \citep{li_strongly_2018, liu_prospects_2019, wucknitz_cosmology_2021}, as well as constraining the fraction of compact objects constituting dark matter \citep{munoz_lensing_2016, sammons_first_2020, connor_stellar_2023}.

Some surveys can now localise FRBs with (sub-)arcsecond accuracy, allowing for the identification of their host galaxies and improving lens parameter estimates, though the detection rates remain low.
Despite the expected low lensing rate of FRB ($10^{-3}-10^{-5}$), the CHIME/FRB experiment is approaching the required detection numbers, although no gravitationally lensed bursts have been identified among the 172 studied so far \citep{leung_constraining_2022}.
Future radio interferometers, such as CHORD, SKA, DSA-2000 and BURSTT, promise to achieve the required FoV, sensitivity, localisation accuracy and spectro-temporal resolution required to find gravitationally lensed FRBs. These advancements present exciting prospects for the development of time-domain gravitational lensing, potentially leading to groundbreaking discoveries in the coming decade. 


\ack{IPM would like to thank Graham Smith, Martin Hendry and Federica Bianco for the opportunity to present this work, as well as Ben Stappers and Andr\'es G\'urpide for their useful input and discussions, and the anonymous referees for their valuable input. IPM further acknowledges funding from an NWO Rubicon Fellowship, project number 019.221EN.019.}

\bibliographystyle{naturemag-doi}
{\small\bibliography{biblio}}

\end{document}